\documentclass{egpubl}
\usepackage{pg2026s}

\WsConferencePaper

\usepackage[T1]{fontenc}
\usepackage{dfadobe}  

\usepackage{cite}  
\BibtexOrBiblatex
\electronicVersion
\PrintedOrElectronic
\ifpdf \usepackage[pdftex]{graphicx} \pdfcompresslevel=9
\else \usepackage[dvips]{graphicx} \fi

\usepackage{egweblnk} 

\usepackage[switch]{lineno}

\usepackage{amsfonts}
\usepackage{algorithm}
\usepackage{algorithmic}
\usepackage{amsmath}
\usepackage{amssymb}
\usepackage{array}
\usepackage{booktabs}
\usepackage{multicol}                 
\usepackage{multirow}          
\usepackage{pifont}
\usepackage{makecell}
\usepackage[caption=false,farskip=0pt,labelfont={bf}]{subfig}
\usepackage[autolanguage]{numprint}

\usepackage{xcolor}
\usepackage{graphicx}

\title
      {Segmenting Geometric Textures of 3D Meshes}

\author[Weidan Xiong, Kaijie Qu, Shihao Feng, Shunqi Fan, Rongyu Xiong, and Hui Huang]
{\parbox{\textwidth}
    {\centering
        Weidan Xiong$^{1}$ \orcid{0000-0001-5490-3676},
        Kaijie Qu$^{1}$ \orcid{0009-0003-1151-0241},
        Shihao Feng$^{1}$ \orcid{0009-0000-7371-2058},   
        Shunqi Fan$^{2}$,
        Rongyu Xiong$^{3}$ \orcid{0009-0008-3234-6803},
        and 
        Hui Huang$^{1}$\thanks{Corresponding author.} \orcid{0000-0003-3212-0544} 
    }
    \\
    {\parbox{\textwidth}
        {\centering 
            $^1$Guangdong Provincial Key Laboratory of Visual Media and Multidimensional Intelligence, CSSE, Shenzhen University, China \\
            $^2$NetEase, China \qquad
            $^3$Nanyang Technological University, Singapore 
        }
    }
}

\begin{document}

\teaser{
\includegraphics[width=\linewidth]{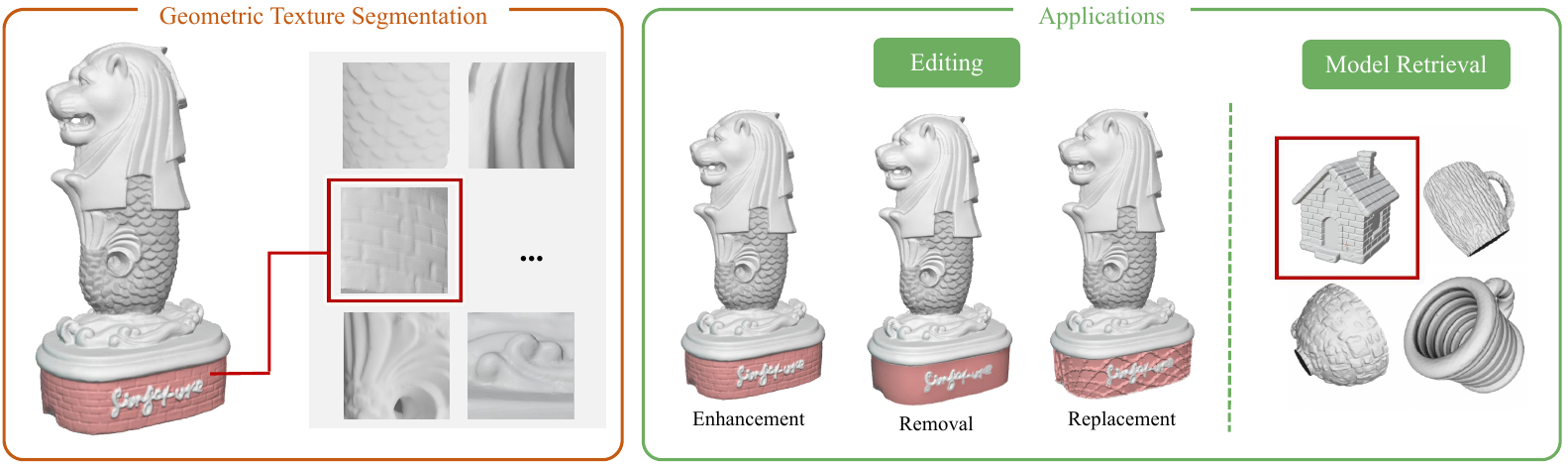}
 \centering
  \caption{
  Our method can segment 3D meshes based on 3D geometric texture, without requiring predefined label sets or precisely tailored text prompts. We develop an interactive tool to support a range of applications, including surface editing and model retrieval.
}
\label{fig:teaser}
}

\maketitle
\begin{abstract}

Recognizing geometric texture patterns on 3D models is critical for applications such as heritage protection, model editing, and 3D printing. Existing 3D model segmentation methods predominantly focus on identifying high-level semantic parts, thereby failing to capture middle-level textures. While recent Vision-Language Models can perform image segmentation with appropriate text prompts, they still struggle to recognize 3D geometric textures. In this work, we present a method that segments 3D mesh models based on middle-level geometric textures. We further introduce a new geometric texture segmentation benchmark and fine-tune SAM3 accordingly. The segmented texture regions are then clustered based on DINO feature similarity and lifted from 2D to 3D. Additionally, we develop an interactive tool to enable both 3D model editing based on our segmentation results and model retrieval from the dataset based on geometric and texture similarity. Experiments demonstrate that our method significantly outperforms existing methods in segmenting geometric textures and can facilitate various applications.



\end{abstract}  

\section{Introduction}
\label{sec:intro}

\begin{figure}
\centering
  \includegraphics[width=\linewidth]{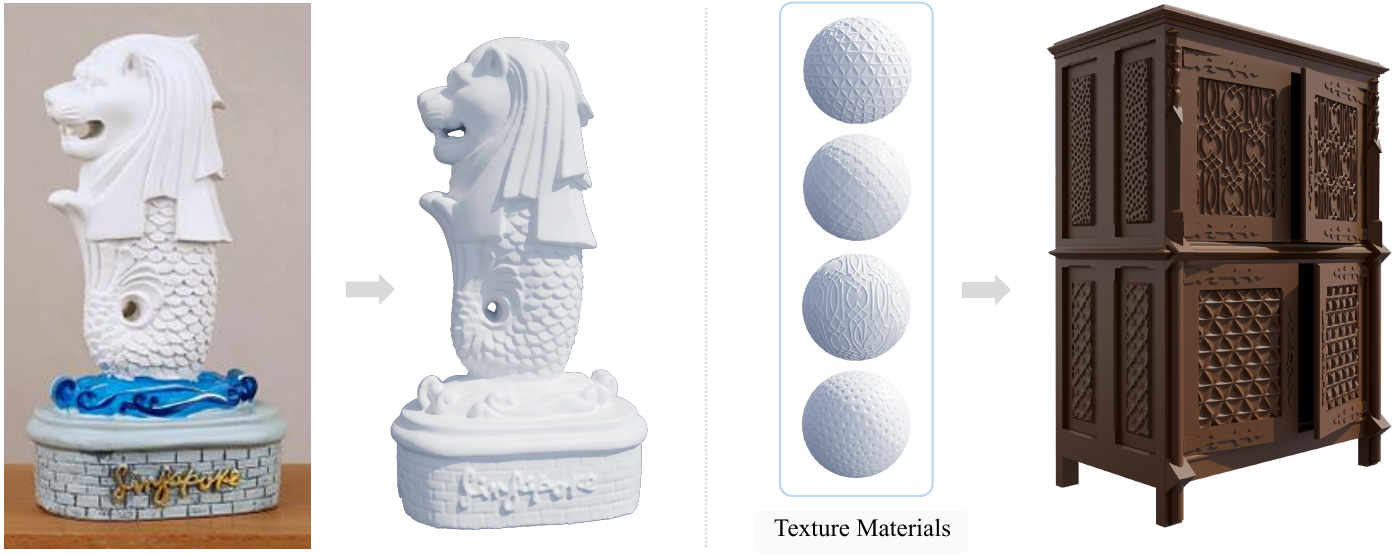}
  \caption{High-quality 3D digital models with intricate geometric textures obtained from scanning (left) and manual sculpting (right).}
  \label{fig:geotex}
\end{figure}

Geometric textures/patterns constitute local, repetitive geometric variations across a 3D surface, remaining independent of the global model structure~\cite{biasotti2018shrec,paolini2025shrec}. High-quality 3D models featuring intricate geometric textures are typically acquired via physical scanning or manual design (see Figure~\ref{fig:geotex}). These assets drive critical downstream applications, ranging from digital twins and heritage preservation to immersive gaming and 3D printing~\cite{xiong2026paver}. Effective recognition of geometric textures is therefore essential, facilitating cultural heritage analysis (e.g., reassembling fragmented artifacts~\cite{biasotti2018shrec}) and enabling texture-aware 3D editing workflows. However, current practice relies heavily on manual inspection and laborious analysis by users or modelers, which is both time-consuming and inefficient.

However, existing model analysis and segmentation methods predominantly operate either at high-level semantics~\cite{hanocka2019meshcnn, Vecchio2023Met} or at low-level geometric features (e.g., curvature, normals, and noise)~\cite{wang2018shape, Guerrero2018PCPNet}, often failing to capture the intricate patterns of middle-level geometric textures. While professional tools like ZBrush facilitate interactive, local modifications, they lack the capability for holistic, global editing of regions characterized by specific texture patterns in a single pass. Consequently, despite growing research interest in recognizing and editing geometric textures on 3D surfaces~\cite{Ganapathi20253dTexSeg}, a robust and comprehensive solution remains an open challenge.

In this work, we propose a framework for geometric texture recognition on 3D surface models. 
Leveraging a 2D-to-3D paradigm, our method begins by rendering multi-view images from the input 3D mesh. 
We then employ a visual segmentation model fine-tuned on our newly constructed benchmark to identify texture regions within the rendered images. 
Corresponding features are extracted from these segmented patches and clustered into semantic groups. 
Subsequently, we introduce a weighted voting strategy to fuse these 2D predictions and lift \emph{them} into a coherent 3D segmentation. 
Finally, we adopt a filter-based geometric editing technique and develop an interactive tool that enables users to perform holistic, one-pass editing of specific texture patterns on the 3D model based on the segmentation results. 
The main contributions are summarized as follows:
\begin{itemize}
    \item A texture material dataset and a geometric texture segmentation benchmark. The datasets are publicly available at: \url{https://github.com/ALingll/GTSeg};
    \item A geometric texture segmentation method for 3D meshes, based on multi-view rendering and visual foundation models;
    \item An interactive, texture-oriented 3D model editing tool, supporting model segmentation and retrieval as well as geometric texture enhancement, removal, and transfer.
\end{itemize}

\section{Related Works}
\label{sec:related-works}

\noindent\textbf{\textit{Geometric Texture}}
over 3D model surfaces are high-frequency, repetitive surface details distinct from global shape and random noise (excluding hollowing and winding structures).
Existing research on geometric textures mainly focused on synthesis and interactive editing.
Early works established procedural and example-based methods for texture synthesis \cite{bhat2004geometric,elber2005geometric,zheng2024design}, a direction recently advanced by deep generative models \cite{hertz2020deep,yang2025genvdm}.
Complementarily, editing tools have evolved from manual manipulation of displacement maps \cite{lai2005geometric} to recent texture generation based on the vision-language model \cite{meng2025text2vdm}.
However, as the prerequisite for automated analysis and editing, the fundamental problem of geometric texture recognition and segmentation is significantly less explored.
Several studies have attempted to bridge this gap. One direction is to separate texture and non-texture regions using learning-based methods~\cite{ganapathi20243d,paolini2024binary}.
Tortorici et al.~\cite{tortorici2021convolution} introduced a convolution-based descriptor for recognizing user-specific textures. 

Despite these efforts, current methods still exhibit poor generalization when dealing with multiple textures, multi-scale structures, and complex topologies. 
To the best of our knowledge, an open-source benchmark, a robust and automated solution for geometric texture segmentation and recognition that handles diverse and complex patterns over a complicated model, is still missing in the literature~\cite{biasotti2018shrec,paolini2025shrec}.

\noindent\textbf{\textit{Model Segmentation}}
is an essential task in computer graphics and vision. 
Early methods mainly rely on low-level geometric primitives such as curvature, surface normals, and geodesic distances to partition 3D models, including region growing~\cite{liang2020survey}, graph cuts~\cite{katz2003hierarchical}, spectral clustering, and MRF/CRF-based optimization~\cite{shamir2008survey}. Although these methods perform stably on segmenting homogeneous low-level geometric regions (e.g., planar surfaces, smooth curved patches), they fail to capture high-frequency, repetitive geometric patterns that define geometric textures.
Recent deep learning works focus on semantic or part-level segmentation. For native 3D representations, point cloud networks (e.g., PointNet~\cite{qi2017pointnet,qi2017pointnetplusplus}, DGCNN\cite{Wang2019DGCNN}) and mesh-based networks~\cite{hanocka2019meshcnn} achieve state-of-the-art performance on benchmarks such as ShapeNet Part\cite{chang2015shapenet} and PartNet\cite{Mo2019PartNet}. However, these methods are limited to coarse-grained category or part labels (e.g., separating chair legs from seats) and cannot recognize fine-grained geometric textures~\cite{yin2024sai3d,xu2025sampro3d,liu2025partfield}. 
Recent works following a 2D-to-3D pipeline, leverage pre-trained 2D foundation models (e.g., SAM~\cite{Kirillov2023SAM}, CLIP~\cite{Chen2025CLIP}) to project multi-view 2D predictions back to 3D~\cite{zhong2024meshsegmenter,Tang2024SAM,sam3dteam2026sam3d,ZHOU2026OV3dSeg}. While these methods enable open-vocabulary segmentation, they still focus on semantic object categories rather than geometric texture patterns.

To the best of our knowledge, the segmentation of geometric textures on 3D models remains an open problem~\cite{paolini2025shrec}. 
Current attempts indicate that image-like local descriptors offer the most promising solution for geometric texture recognition and segmentation~\cite{biasotti2018shrec}. 
Hence, we borrow the ability from the visual foundation model and employ the 2D-to-3D pipeline to address this problem.

\noindent\textbf{\textit{Image Segmentation}}
aims to partition an image into regions that are coherent in semantics, instance identity, or visual attributes. Earlier deep segmentation methods primarily focus on dense prediction and instance mask modeling \cite{long2015fcn, ronneberger2015unet,he2017maskrcnn}.
Recent segmentation research has increasingly shifted toward unified and promptable foundation models. MaskFormer and Mask2Former cast semantic, instance, and panoptic segmentation as mask classification \cite{cheng2021maskformer,cheng2022mask2former}. 
SAM3 introduces strong promptable concept segmentation, enabling the detection and segmentation of all instances matching a text or exemplar concept \cite{carion2025sam3}.
SAMesh lifts multi-view SAM2 masks onto mesh surfaces for zero-shot mesh part segmentation \cite{Tang2024SAM}. These studies suggest that SAM-family models are effective front-ends for 3D model segmentation.
Recently, textureSAM validates the zero-shot capability of SAM in 2D texture segmentation after fine-tuning with sufficient training data \cite{cohen2026decoupling}. 

However, unlike the 2D textures provided by TextureSAM, the identical geometric texture can look dramatically different across viewpoints, which implies the critical role of proper training data and geometric constraints.
Therefore, in this work, we do not directly treat 2D masks as the final predictions; instead, we leverage both image and geometry evidence, combining curvature re-rendering, mask-level texture-feature clustering, and multi-view 3D voting to obtain view-consistent geometric texture segmentation.

\section{Geometric Texture Segmentation Benchmark}
\label{subsec:dataset}

\begin{figure}
    \centering
    \includegraphics[width=\linewidth]{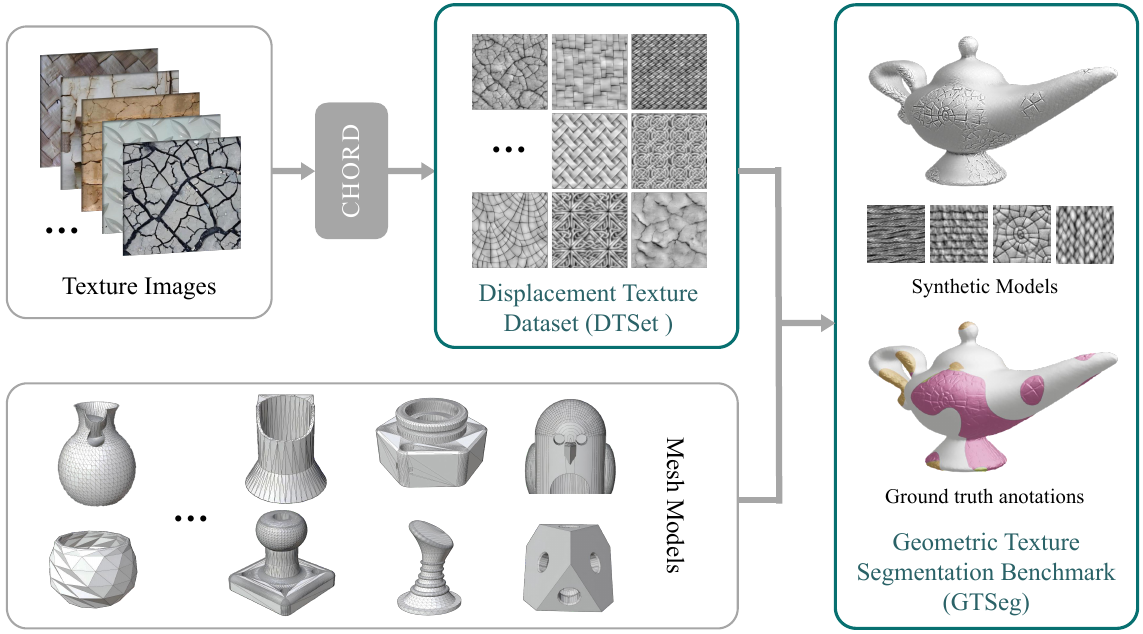}
    \caption{The generation pipeline of our DTSet dataset and GTSeg benchmark.}
    \label{fig:benchmark}
\end{figure}

To begin, we construct a large-scale geometric texture segmentation benchmark, consisting of synthetic and annotated 3D models.
Collecting reconstructed models through scanning is extremely expensive and labor-intensive. Instead, we choose to sculpt 3D models via texture displacement maps, see Figure~\ref{fig:benchmark}. 

\noindent\textit{Texture Displacement Maps Dataset.}
We use Chord~\cite{ying2025chord} to generate displacement maps based on the open-source 2D texture images dataset DTD~\cite{cimpoi2014describing}. 
We construct a displacement map dataset, \textit{TDSet}, consisting of 350 diverse displacement maps. 
The displacement maps in TDSet are employed to synthesize 3D geometric textures over 3D models.

\noindent\textit{3D Geometric Texture Benchmark.}
Next, we construct a geometric texture benchmark called \textit{GTSeg}, which consists of a training set and a test set. For the training set, we collect 80 synthetic models from Thingi10K~\cite{Zhou2016Thingi10K}. We randomly select four or five displacement maps from TDSet and apply them to the surface of each model. All geometric texture instances are automatically annotated. For each model, we repeat this process ten times. The final training set consists of 800 models, each with multiple geometric texture categories and instances. Moreover, we construct a test set comprising 20 synthetic models generated following the above procedure (with no overlap with the training set) and 10 real scans from \cite{xiong2014robust} and public domain. We manually annotate the geometric texture instances on the real scans.

\begin{figure*}[t]
    \centering
    \includegraphics[width=\textwidth]{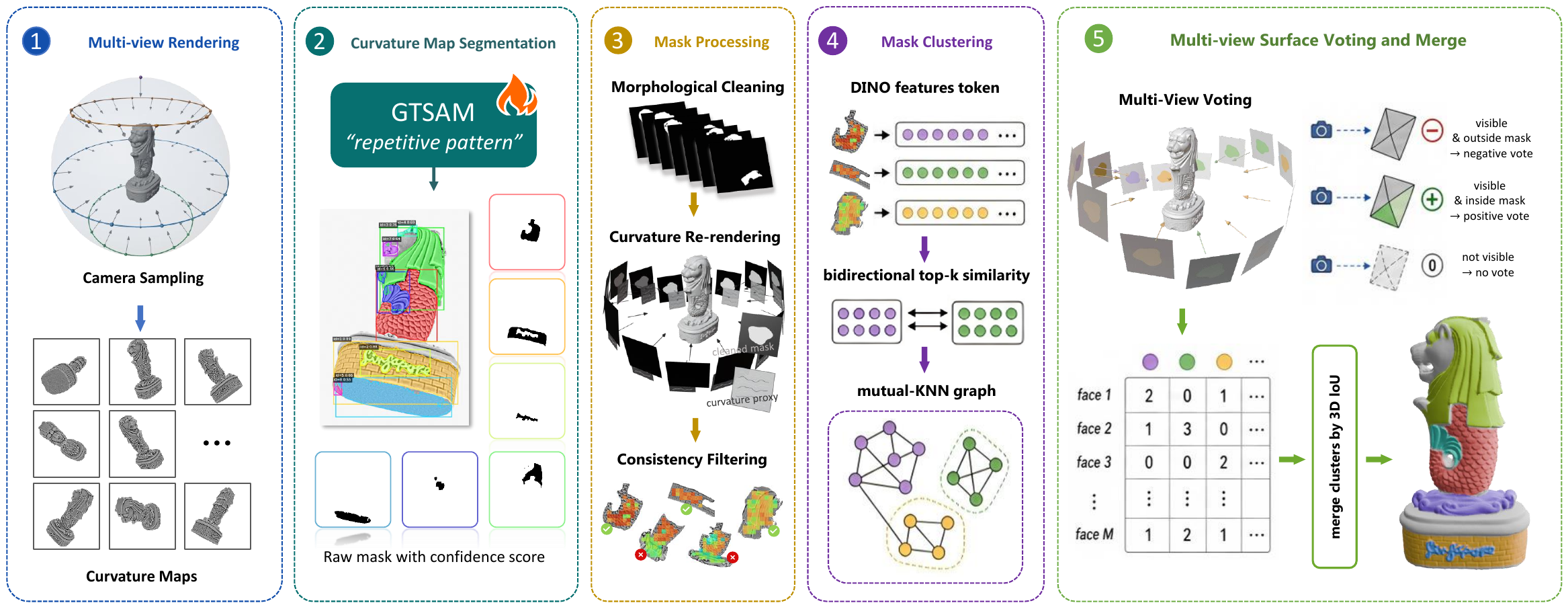}
    \caption{Overview of our 3D geometric texture segmentation framework.}
    \label{fig:framework}
\end{figure*}

\section{Geometric Texture-oriented Image Segmentation Model}
\label{sec:sam3}

In this section, we develop a 2D image segmentation model specifically for segmenting 3D geometric textures. We first justify our choice of representation, and subsequently detail the training dataset construction pipeline and the model fine-tuning protocol.

Although recent advancements have made significant progress in texture-aware 2D image segmentation, they cannot be directly generalized to 3D geometric texture recognition. TextureSAM~\cite{cohen2026decoupling} mitigates the semantic bias of foundation models by fine-tuning on 2D RGB images where object regions are synthetically replaced with descriptive visual textures. In contrast, 3D geometric textures lack explicit colors and depend solely on local topographic variations. Furthermore, the 2D projection of the same 3D geometric pattern exhibits severe perspective distortions and foreshortening across different viewing angles. Consequently, without explicit adaptation to intrinsic 3D geometry and multi-view consistency, models constrained to 2D visual priors inevitably fail to achieve view-consistent segmentation of complex 3D surface patterns.

\noindent\textit{Grayscale Curvature and Mask Rendering.}
Surface curvature is an intrinsic geometric property that remains strictly invariant to lighting conditions and rigid transformations. 
To bridge 3D geometry with 2D segmentation, we compute per-vertex mean curvature as the raw scalar $s_i$.
To mitigate the impact of mesh singularities and ensure high image contrast, we clip outliers according to the 5th/95th percentiles. The curvature values are then normalized to [0,1] via min-max normalization.

Here, we split the models in the training set of GTSeg into two sets: 76 models are used for training and 4 for validation. We rendered each model to generate curvature maps and ground-truth masks from 14 uniformly sampled viewpoints. 
That is to say, the training set contains 10,640 image and mask pairs, while the validation set contains 560 image and mask pairs. All rendered views and synthesized variants derived from the same source model are assigned to the same split to prevent data leakage.
Finally, we generate a dataset of 11,200 pairs of view-dependent curvature maps and corresponding ground-truth masks. The training set will be employed to adapt the vision foundation model to our task later.

\noindent\textit{Model Fine-tuning.}
While the pre-trained SAM3 model demonstrates strong zero-shot capabilities in general concept segmentation, it struggles to precisely delineate non-semantic, mid-level geometric textures. To efficiently bridge this domain gap without catastrophic forgetting, we employ Low-Rank Adaptation (LoRA)~\cite{hu2022lora} to fine-tune the model on our synthesized dataset and obtain GTSAM.

The recognition of geometric textures relies heavily on mid-level patterns rather than high-level linguistic concepts. Hence, we freeze the entire language encoder to preserve general semantic priors and reduce computational overhead. To this end, we inject trainable rank-decomposition matrices into the Vision Encoder, Geometry Encoder, and the decoding modules. This targeted injection maximizes the model's sensitivity to high-frequency textural signals while maintaining its robust zero-shot foundation.

\section{Geometric Texture-oriented Model Segmentation}

In this section, we introduce a framework to address the problem of geometric texture segmentation, via formulating it as a task of multi-view consistency perception. The overview of our framework is visualized in Figure~\ref{fig:framework}.

Given a 3D mesh $\mathcal{M}$, we first render a set of grayscale curvature maps $\{C_i\}_{i=1}^N$ from a collection of predefined camera poses $\{P_i\}_{i=1}^N$.
Subsequently, the rendered images are fed to our GTSAM to predict a set of pixel-level geometric texture masks. We introduce a feature association module based on DINOv2 to maintain semantic consistency across disparate views. This module extracts and clusters high-dimensional feature descriptors from each mask, thereby establishing label correspondences across different viewpoints in a unified latent space. 
Finally, multi-view masks with consistent labels are projected onto the model surface $\mathcal{M}$. To resolve prediction conflicts arising from occlusions or perspective distortions, we implement a spatial-consistency weighted voting mechanism. This mechanism performs per-face arbitration to make definitive label decisions, yielding precise and smooth geometric texture segmentation results.

\subsection{Texture-oriented Image Segmentation}
We set $N$ cameras to render the input mesh, ensuring complete coverage and sufficient overlap. In our implementation, we let $N=50$.
The $t$-th camera renders a curvature image $I_t$.
Next, we send the rendered curvature images to our GTSAM. 
For $I_t$, GTSAM predicts a set of $N_t$ instance masks. Each instance mask $M_{t,m}$ is paired with a confidence score $c_{t,m}\in[0,1]$.

For each predicted mask $M_{t,m}$, we perform 8-connected component analysis. 
Components with an area smaller than the threshold $A_{\min}$ are discarded.
We retain those sufficiently large connected components
and treat them as independent mask instances. This prevents disconnected 2D regions from being merged into the same texture cluster.

\subsection{Feature-based Patch Clustering.}

Although GTSAM produces high-quality class-agnostic masks $\{M_{t,m}\}$, these masks are generated independently for each view and lack explicit semantic correspondence. To assign accurate instance labels to disparate views, we leverage DINOv2 as a robust feature extractor. It's capable of achieving cross-view semantic consistency in the absence of per-view labels~\cite{duggal2025eval3d}.

\noindent\textit{Region Re-rendering.}
After that, we project each mask component onto the mesh.
The set of faces covered by the mask component $C_a$ is denoted as $P_a$.
We discard mask components that contain very few faces and store the remaining components as a group. 

To extract a view-normalized texture descriptor, we re-render each projected region. The region center and average normal are estimated as:
\begin{equation}
    \mathbf{o}_a = \frac{1}{|\mathcal{V}(P_a)|} \sum_{\mathbf{v}_i\in\mathcal{V}(P_a)} \mathbf{v}_i,
    \qquad
    \mathbf{n}_a = \frac{\sum_{\mathbf{v}_i\in\mathcal{V}(P_a)}\mathbf{n}_i}{\left\| \sum_{\mathbf{v}_i\in\mathcal{V}(P_a)}\mathbf{n}_i \right\|_2}.
\end{equation}

An orthographic camera is then placed along $\mathbf{n}_a$ and oriented toward $\mathbf{o}_a$, producing a local curvature rendering $J_a$ for the mask instance.

Next, we extract the DINOv2 patch tokens of local curvature rendering $J_a$~\cite{oquab2024dinov2}. The visible region is cropped and resized to a fixed resolution. Each rendered view is augmented with rotations ($0^\circ$, $90^\circ$, $180^\circ$, and $270^\circ$) and horizontal flip. We then generate patch tokens using DINOv2 for each augmented view. The region mask is down-sampled to the DINO patch grid, and the tokens covered by the mask are sampled to create a token pool:
$\mathcal{T}_a = \{\hat{\mathbf{z}}_{a,1}, \ldots,\hat{\mathbf{z}}_{a,K_a}\}$.
Each token is normalized into a unit vector.

\begin{figure}[t]
    \centering
    \includegraphics[width=\linewidth]{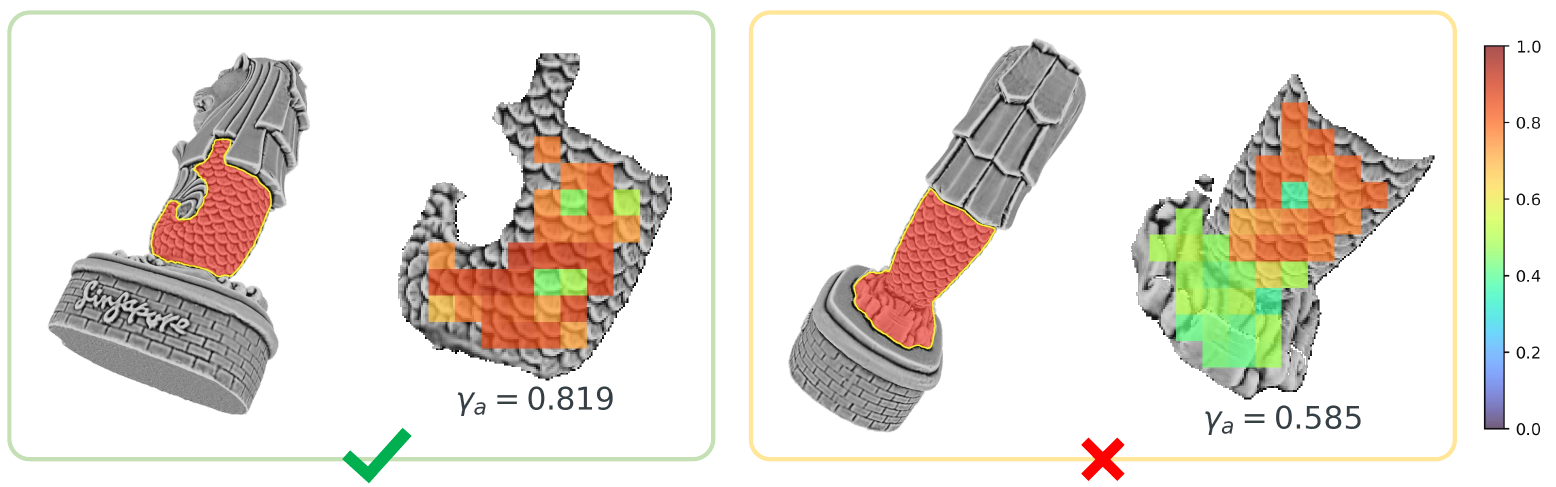}
    \caption{
    The intra-mask consistency on two masks of the Merlion example. 
    For each view, we visualize the curvature map (left) with its candidate mask marked in red. The token-to-centroid cosine similarity map is provided on the right. The value of cosine similarity is visualized from red (high) to blue (low).
    }
    \label{fig:mask_consistency_filtering}
\end{figure}

\noindent\textit{Consistency Filtering.}
To eliminate masks that contain inconsistent textures, we compute a consistency score $\gamma_a$ for each $J_a$. This score is defined as the median similarity between each local feature $\hat{\mathbf{z}}_{a, i}$ and the normalized average feature direction of the entire mask. The score is formulated as:
\begin{equation}
    \gamma_a = \operatorname{median}_i \left( \frac{\hat{\mathbf{z}}_{a,i}^{\top} \sum_j \hat{\mathbf{z}}_{a,j}}{\left\|\sum_j \hat{\mathbf{z}}_{a,j}\right\|_2} \right).
\end{equation}

We eliminate masks with $\gamma_a$ smaller than a predefined threshold $\tau_s$. In our implementation, we let $\tau_{s}=0.65$. 
Two masks of the Merlion example are provided in Figure~\ref{fig:mask_consistency_filtering}. The mask on the left contains only one geometry texture, therefore exhibits high consistency. Meanwhile, the mask on the right is eliminated since it contains two types of texture, which deliver a consistency score smaller than the threshold.

\begin{figure}[t]
    \centering
    \includegraphics[width=0.85\linewidth]{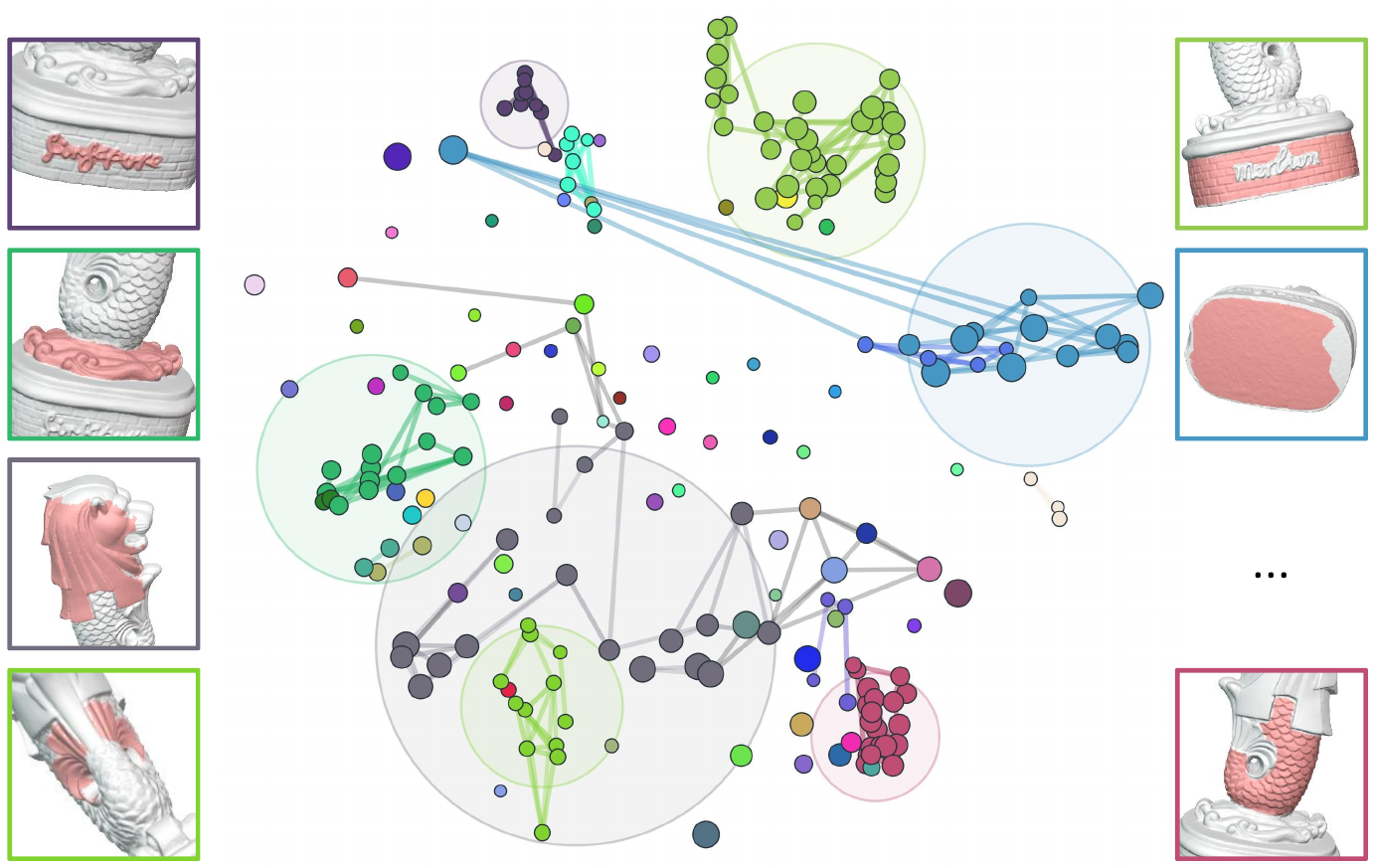}
    \caption{
    Visualization of mask clustering in the token-distance embedding space.
    Nodes denote mask instances, with sizes proportional to mask areas and colors indicating final clusters.
    Positions are obtained by metric MDS~\cite{borg2005modern} from the pairwise distance: $1-\operatorname{sim}(a,b)$.
    Edges represent valid mutual-$k$NN connections, with thicker edges indicating higher similarity.
    }
    \label{fig:clustering}
\end{figure}

\noindent\textit{Feature Clustering.}
To merge instances representing the same geometric texture, for each pair of mask instances $J_a$ and $J_b$, we compute a token-token similarity matrix $\mathbf{S}^{ab}_{ij} = \hat{\mathbf{z}}_{a,i}^{\top} \hat{\mathbf{z}}_{b,j}$. 
We utilize bidirectional top-$k$ matching to reduce the sensitivity to boundary tokens and partial occlusions. 
Specifically, we first identify the maximum matching scores along both directions: $u_i=\max_j \mathbf{S}^{ab}_{ij}$ and $v_j=\max_i \mathbf{S}^{ab}_{ij}$. The pairwise instance similarity is then defined as the average of the top-$k$ mean values from both sets:
\begin{equation}
    \operatorname{sim}(a,b)
    =
    \frac{1}{2}
    \left[
    \operatorname{TopKMean}_k(\{u_i\}_i)
    +
    \operatorname{TopKMean}_k(\{v_j\}_j)
    \right].
\end{equation}

To segment geometric texture instances, we not only consider the similarity of the rendered textures, but also consider their connectivity over the surface. We store the disconnected regions with the same pattern as different texture instances.
We first construct a mutual-$k$NN graph $\mathcal{G}$ over all valid mask instances~\cite{yan2024maskclustering}.
The nearest neighbors $k$ of $a$ are denoted as $\mathcal{N}_k(a)$. We create an edge between $a$ and $b$ if they are mutual neighbors ($b\in\mathcal{N}_k(a)$ and $a\in\mathcal{N}_k(b)$) and their similarity satisfies $\operatorname{sim}(a,b)\ge \tau_{\mathrm{clu}}$. Subsequently, the raw mask clusters $\{\mathcal{Q}_r\}$ are obtained by extracting the connected components of $\mathcal{G}$, formulated as $\{\mathcal{Q}_r\} = \operatorname{CC}(\mathcal{G})$.

To refine the clusters, we calculate a cluster support score for each mask $a \in \mathcal{Q}$, defined as the median similarity to its cluster peers: $\rho_a = \operatorname{median}_{b\in\mathcal{Q}, b\ne a} \operatorname{sim}(a,b)$. Low-support members are split into singleton clusters. Finally, the mask confidence used for 3D voting is reweighted by incorporating its internal consistency $\gamma_a$ and cluster support $\rho_a$:
\begin{equation}
    \hat{s}_a
    =
    \max
    \left(
    s_{\min},
    s_a \gamma_a^{\alpha}\rho_a^{\beta}
    \right).
\end{equation}

In our implementation, we let $k=8$ for the bidirectional top-$k$ matching, $k=5$ for the mutual-$k$NN graph construction, and $\tau_{\mathrm{clu}}=0.8$. For the confidence reweighting, we set $\alpha=1.0$, $\beta=1$, and $s_{\min}=0.25$. A visualization of the resulting mask clusters is shown in Figure~\ref{fig:clustering}.

\subsection{Mask Fusion by 3D Voting}
To fuse the 2D mask clusters $\{\mathcal{Q}_r\}$ into the 3D mesh, we project the masks onto the mesh surface and accumulate votes on the visible faces. For a mask $C_a \in \mathcal{Q}_r$ and a view $t_a$, a mesh face $f$ will receive a positive weight $w_+$ when its projected face center falls inside $C_a$, and will receive a negative penalty $-w_-$ otherwise. This vote is further scaled by the mask confidence score $cs_a$ and an angular weight $\cos(\theta_{a,f})$, where $\theta_{a,f}$ is the angle between the face normal and the viewing direction.
The raw voting score is accumulated as:
\begin{equation}
    V_{r,f}^{\mathrm{raw}}
    =
    \sum_{a \in \mathcal{V}_{r,f}}
    c_a \cos(\theta_{a,f}) \cdot W(a, f),
\end{equation}
where $\mathcal{V}_{r,f} \subseteq \mathcal{Q}_r$ denote the subset of masks from which face $f$ is visible, $W(a, f) = w_+$ if the projected face center of $f$ lies inside mask $C_a$, and $-w_-$ otherwise. To avoid bias toward frequently observed regions, the raw score is averaged by the visibility count, yielding the final score:
\begin{equation}
V_{r,f} = \frac{V_{r,f}^{\mathrm{raw}}}{\max(1, |\mathcal{V}_{r,f}|)}.
\end{equation}

A face is assigned to cluster $r$ if its average score $V_{r,f} > \tau_V$ and it accumulates at least $n_{\min}^{+}$ positive hits. 
Face-label conflicts are resolved by weighted voting. When a face receives positive votes from multiple texture clusters, the cluster with the highest accumulated normalized score is selected.

Finally, we refine the assignments by extracting the connected components over the 3D mesh and discarding isolated components whose area is smaller than a minimum value. Ultimately, we get a set of 3D instances $\{\mathcal{P}_r\}$ from $\{Q_r\}$. In our implement, we set $w_+=2.0$, $w_-=1.0$ and $n_{\min}^{+}=3$.

Finally, different initial 3D parts may correspond to the same physical region due to different perspectives. We therefore merge parts that exhibit strong spatial overlap. For any two 3D semantic parts $\mathcal{P}_a$ and $\mathcal{P}_b$, we calculate their 3D Intersection-over-Union (IoU) as $\operatorname{IoU}(a,b) = |\mathcal{P}_a \cap \mathcal{P}_b| / |\mathcal{P}_a \cup \mathcal{P}_b|$, where $|\cdot|$ denotes the number of mesh faces.

If $\operatorname{IoU}(a,b) \ge \tau_{\mathrm{iou}}$, the two parts are connected in a region-adjacency graph. We set a threshold of $\tau_{\mathrm{iou}}=0.8$ to ensure that only highly overlapping parts are merged. The final non-overlapping 3D regions, denoted as $\{\mathcal{P}_k\}_{k=1}^{K}$, are obtained by extracting the connected components of this graph. 

The first row of Figure~\ref{fig:results-1} shows the final result on the Merlion example. The resulting face regions are rendered with distinct colors. The result indicates that our method effectively recognizes all the distinct geometric textures over the model. Even the text on the base is successfully segmented.

\section{Applications}

\noindent\textit{Geometric Texture Enhancement and Removal.}
For model editing, normal filters are capable of removing all the high-frequency local characteristics, such as noises~\cite{zhang2015guided,xiong2018shape}. Specifically, the static/dynamic filter effectively identifies all the relief structures over the model surface, without recognizing their higher-level patterns.
Based on our segmentation results, we employ the static/dynamic filter to interactively enhance or remove a specific type of geometric texture over the model (see Figure~\ref{fig:teaser}). Our segmentation method dramatically improves editing efficiency.

\noindent\textit{Geometric Texture Replacement.}
After removing the user-specific geometric texture instance, the user can select an arbitrary texture from our DTSet, and apply it to the target region to achieve replacement of geometric texture (see Figure~\ref{fig:teaser}).

\noindent\textit{Model Retrieval.}
Effective retrieval of the reconstructed model that shares similar geometric texture is critical for the projection and restoration of the heritage.
The query in Figure~\ref{fig:teaser} is the brick wall texture of Merlion. The retrieval is performed among four models. For each model, we performed segmentation and extracted the DINOv2 feature tokens of all the texture instances. Then, we measure the similarity between all the texture instances and the target (Eq. 3), returning the model that contains the texture instance with the highest similarity.

\section{Experimental Results}
\label{sec:results}

\subsection{Ablation Study}

Firstly, we evaluate the contribution of three key components.
The LoRA-adapted SAM3 is replaced with the original zero-shot model. Next, the step of feature-based association is removed from the framework. All the masks are grouped solely according to the overlap of their projected 3D regions. Then, feature-based association is removed from the framework, and masks are grouped solely according to the overlap of their projected 3D regions.
The experimental results are shown in Table~\ref{tab:ablation_study}. For all quantitative evaluations, the results are presented as percentages. 

\begin{table}
    \centering
    \caption{
        Ablation study of our instance segmentation pipeline.
        All variants are evaluated using class-agnostic, face-level metrics,
        and all values are reported as percentages.
    }
    \label{tab:ablation_study}
    \resizebox{\linewidth}{!}{
    \begin{tabular}{lccccc}
        \toprule
        Method
        & MUCov $\uparrow$
        & MWCov $\uparrow$
        & SQ$_{50}$ $\uparrow$
        & RQ$_{50}$ $\uparrow$
        & PQ$_{50}$ $\uparrow$ \\
        \midrule
        w/o SAM3 Fine-tuning
        & 35.64 & 49.04 & 81.66 & 34.57 & 29.15 \\

        w/o Clustering
        & 39.91 & 54.26 & 77.96 & 41.03 & 32.79 \\

        w/o Filtering
        & 54.33 & 71.08 & 79.18 & 59.87 & 48.55 \\

        Ours
        & \textbf{61.78}
        & \textbf{78.24}
        & \textbf{83.15}
        & \textbf{70.99}
        & \textbf{59.32} \\
        \bottomrule
    \end{tabular}
    }
\end{table}

The results indicate that replacing the fine-tuned model with zero-shot SAM3 brings the largest degradation in segmentation accuracy. 
Removing feature-based clustering also substantially reduces performance. 
Consistency filtering provides a further improvement over the unfiltered variant. 
The complete method consistently achieves the best performance across all metrics. Overall, the results demonstrate that all three components are necessary for accurate and reliable 3D geometric-texture instance segmentation. 

\begin{figure}
\centering
  \includegraphics[width=0.85\linewidth]{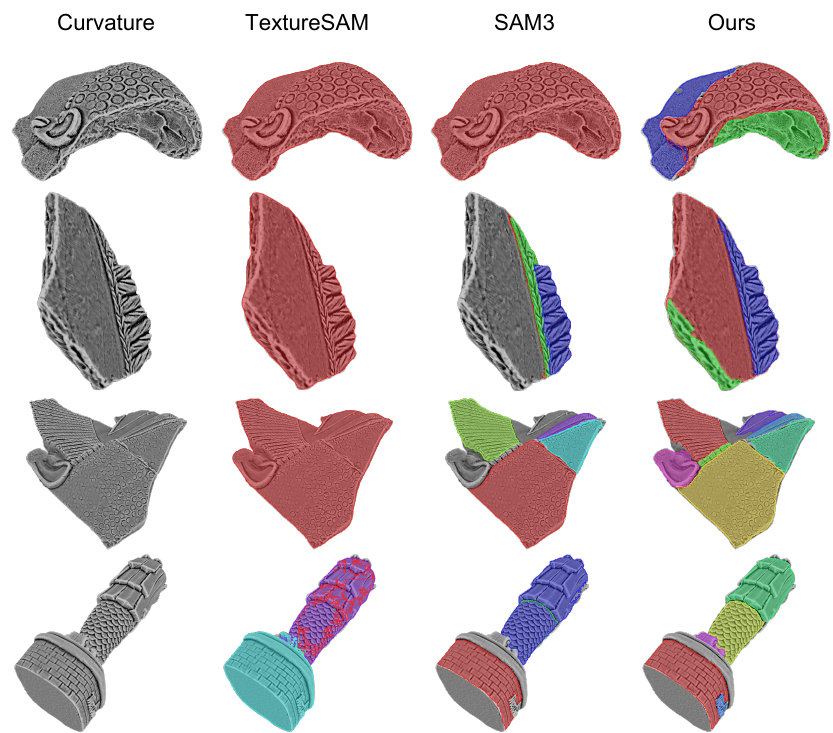}
  \caption{Comparisons of the segmentation results with TextureSAM, SAM3, and our GTSAM on five rendered views of reconstructed real objects.}
  \label{fig:comparisons}
\end{figure}

\subsection{Comparisons on Image Segmentation}

We compare the results of segmentation on rendered curvature maps with zero-shot SAM3~\cite{carion2025sam3} and TextureSAM~\cite{cohen2026decoupling}. 
SAM3 and our methods are given the identical prompt: “\texttt{repetitive pattern}”, in all experiments.

Figure ~\ref{fig:comparisons} qualitatively illustrates the performance of our method compared to the methods described above in wild examples.
We performed quantitative evaluation on the GTSeg test set.
Statistics are provided in Table~\ref{tab:qualitative-comparison}. To comprehensively assess both semantic and instance-level quality, we adopt mIoU and Dice coefficient to measure overall spatial overlap, Precision and Recall for pixel-level exactness, and Average Precision ($\mathrm{AP}$, $\mathrm{AP}_{50}$, $\mathrm{AP}_{75}$) for instance-level localization accuracy.

\begin{table}
\centering
\caption{Quantitative comparison on segmentation of 2D images.}
\label{tab:qualitative-comparison}
\resizebox{\columnwidth}{!}{
\begin{tabular}{lccccccc}
\toprule
Method & mIoU $\uparrow$ & Dice $\uparrow$ & Precision $\uparrow$ & Recall$_{50}$ $\uparrow$ & AP$_{50}$ $\uparrow$ & AP$_{75}$ $\uparrow$ & Mean AP $\uparrow$ \\
\midrule
TextureSAM & 15.61 & 25.87 & 15.67 & 5.73 & 0.70 & 0.17 & 0.43 \\
SAM3 & 59.33 & 69.03 & 72.62 & 36.93 & 29.76 & 17.03 & 23.40 \\
Ours (GTSAM) & \textbf{78.18} & \textbf{85.74} &\textbf{88.90} & \textbf{81.42} & \textbf{74.50} &\textbf{52.72} &\textbf{62.06} \\
\bottomrule
\end{tabular}
}
\end{table}

As shown in Table~\ref{tab:qualitative-comparison}, our fine-tuned SAM3 significantly outperforms all baseline methods on every metric. Specifically, while TextureSAM achieves a high Recall (0.913), its excessively low Precision (0.129) and mIoU (0.128) indicate severe over-segmentation and massive false positives, rendering its instance-level AP scores negligible. The SAM3 demonstrates solid foundational capability with an mIoU of 0.669; however, its performance degrades drastically under stricter boundary criteria ($\mathrm{AP}_{50}$ drops to 0.124), indicating its struggle with precise boundary adherence required by the GTSeg domain. In contrast, our fine-tuned SAM3 effectively bridges this domain gap, achieving a state-of-the-art mIoU of 0.933 and a Dice score of 0.965. Most notably, our method maintains exceptional accuracy under strict intersection thresholds, boosting the Mean AP from 0.218 to 0.890 and $\mathrm{AP}_{75}$ from 0.124 to 0.831. These results demonstrate that our fine-tuning strategy retains the strong recall of the foundation model while drastically improving its precision and boundary localization.

\subsection{Comparisons on Model Segmentation}

We collect 12 models reconstructed from real objects from public resources~\cite{binninger2025tetweave} and synthetic models from SHREC'25 examples ~\cite{paolini2024binary}, with various resolutions, functions, complexities, and semantics. Their geometric textures cover diverse scales, structures, and shapes with open boundaries.

\begin{figure}
\centering
  \includegraphics[width=.93\linewidth]{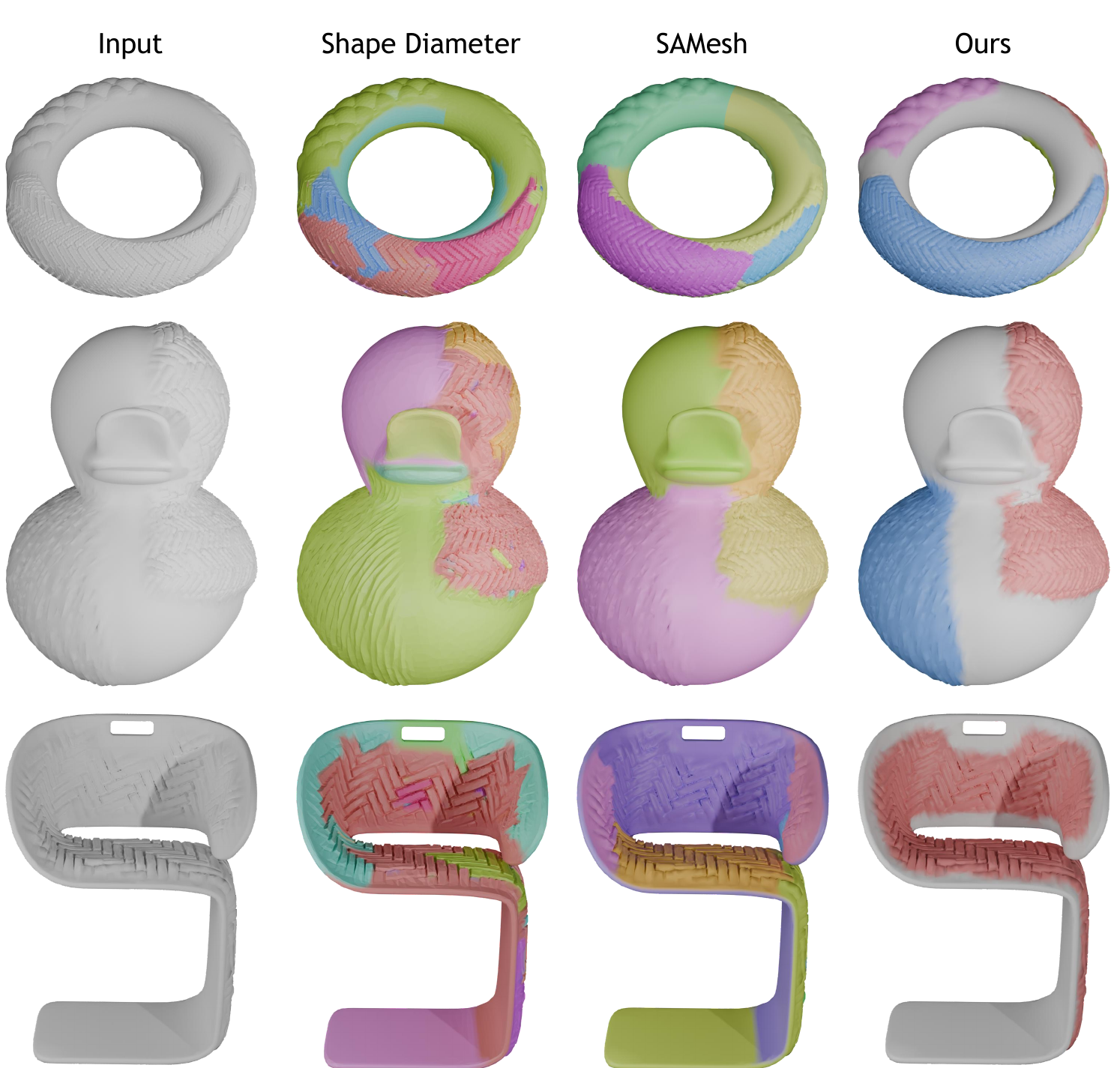}
  \caption{Comparisons of the segmentation results with Shape Diameter, SAMesh, and our method on synthetic 3D models from SHREC'25~\cite{paolini2024binary}.}
  \label{fig:result-synthetic-2}
\end{figure}

\begin{figure*}
\centering
  \includegraphics[width=0.87\linewidth]{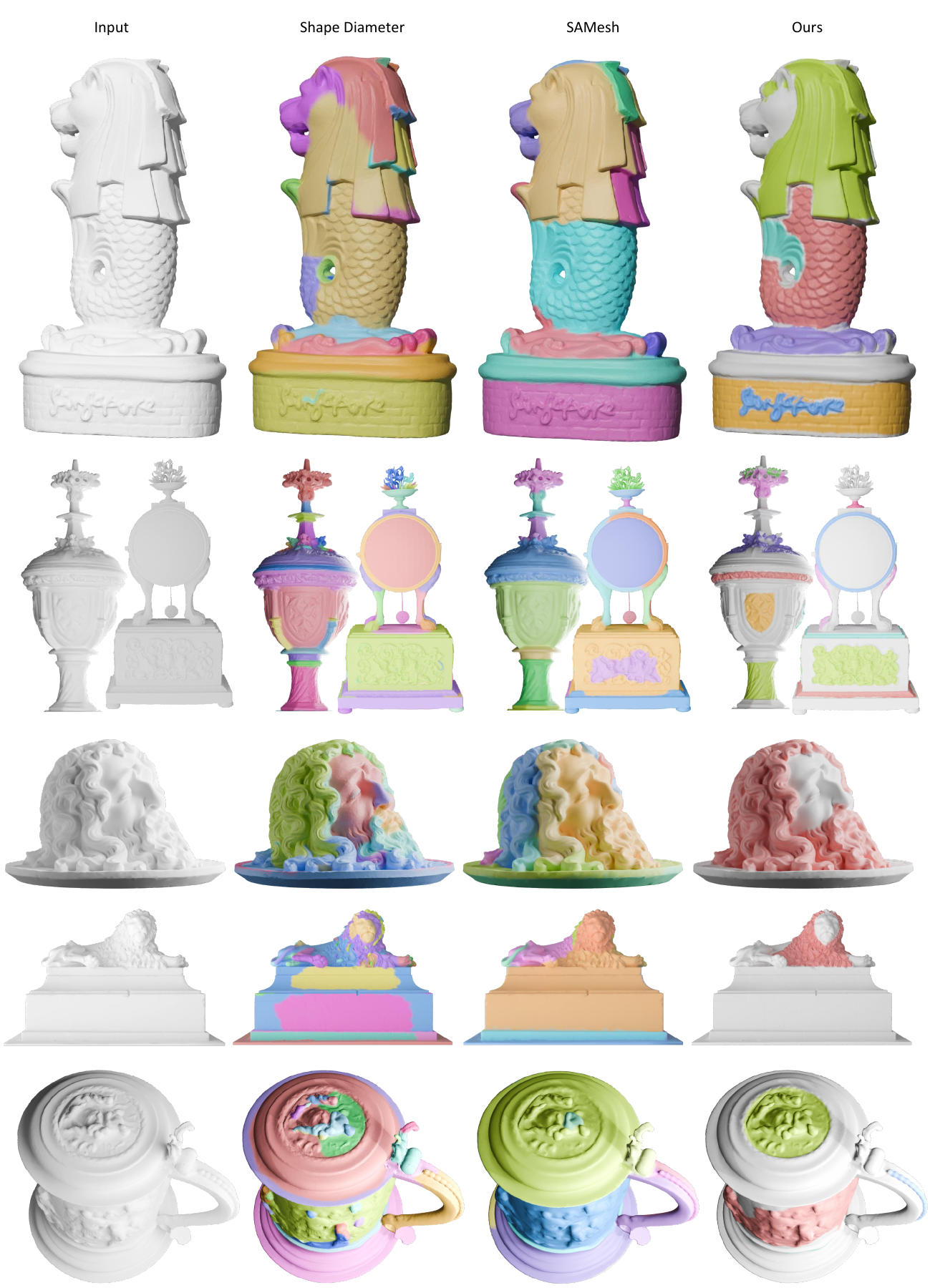}
  \caption{
  Comparisons of the segmentation results with Shape Diameter, SAMesn, and our method on reconstructed real objects. Colors of instances are assigned randomly.}
  \label{fig:results-1}
\end{figure*}

\begin{figure*}
\centering
  \includegraphics[width=0.87\linewidth]{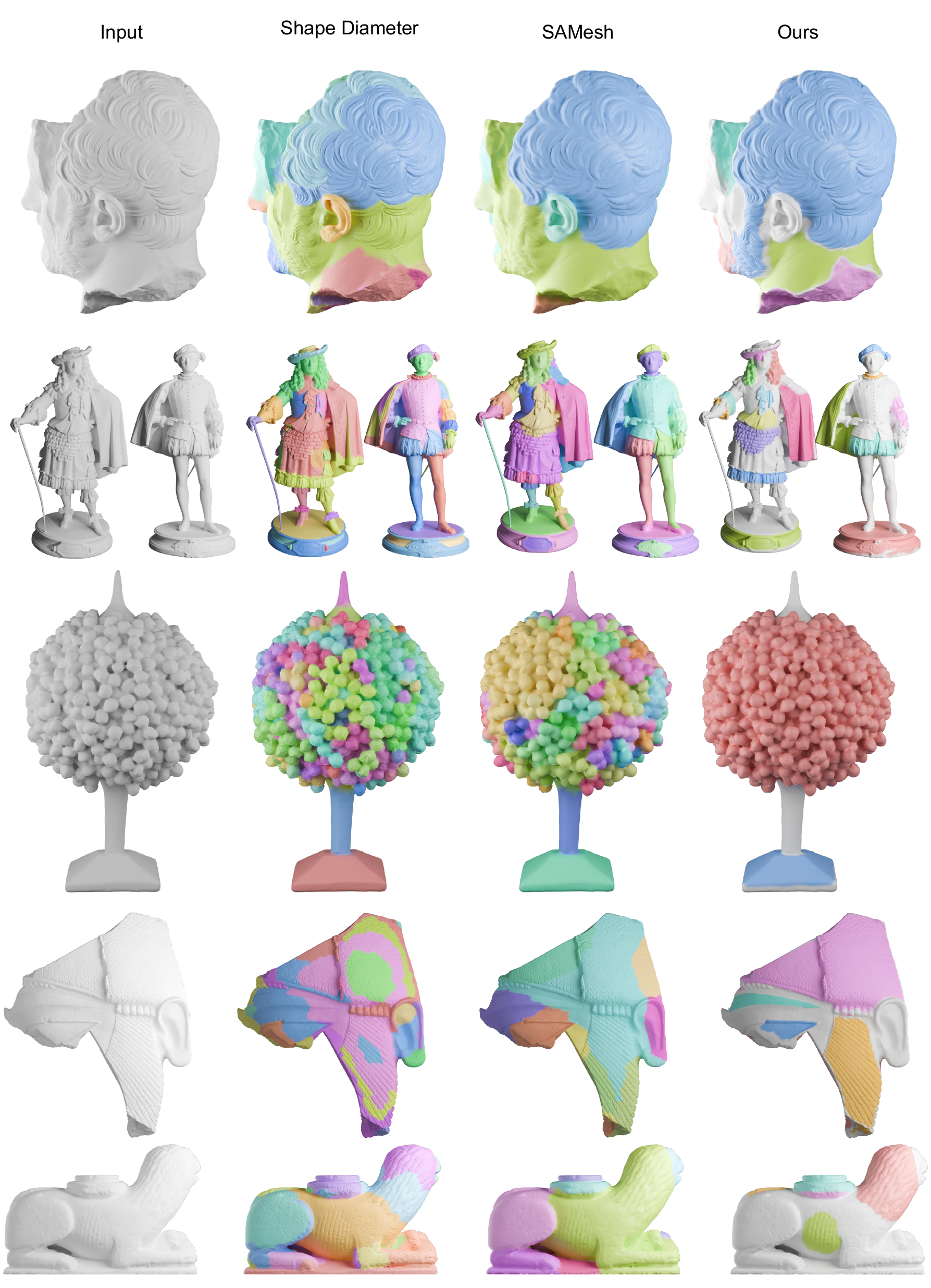}
  \caption{
  Comparisons of the segmentation results with Shape Diameter, SAMesh, and our method on reconstructed real objects. Colors of instances are assigned randomly.}
  \label{fig:results-2}
\end{figure*}

\noindent\textit{{Qualitative Evaluation.}}
We compare our segmentation results with ShapeDiameter~\cite{shapira2008consistent} and SAMesh~\cite{tang2024segment}. 
Figure~\ref{fig:result-synthetic-2} shows a qualitative comparison of three unlabeled models from SHREC'25. Figure~\ref{fig:results-1} and Figure~\ref{fig:results-2} further compare the performance of these methods on real scans.

ShapeDiameter relies on volumetric thickness and global shape structure, making it insensitive to fine-grained surface-level texture boundaries, particularly when textured patterns are shallow, repetitive, or confined to the same semantic part. SAMesh offers better flexibility by projecting 2D segmentation onto 3D meshes, but its behavior is dominated by object/part-level visual priors, leading to cross-view inconsistencies when texture elements share similar semantic appearance, yet differ geometrically. Our method addresses these limitations by using curvature renderings as input, directing segmentation toward intrinsic surface variation rather than color, illumination, or semantic category. Combined with mask-level clustering and multi-view voting, it produces coherent, view-consistent regions for repeated geometric patterns and resolves texture boundaries defined by subtle local surface relief. These comparisons confirm that conventional mesh segmentation and general 2D-to-3D pipelines are insufficient for geometric texture segmentation, and that curvature-based representation with texture-aware clustering provides a more suitable formulation.

\begin{table}
\centering
\caption{Quantitative comparison on segmentation of 3D models.}
\label{tab:quantitative-comparison-model}
\begin{tabular}{lcc}
\toprule
Method & MUCoV $\uparrow$ & Recall$_{50}$ $\uparrow$ \\
\midrule
TextureSAM + Voting & 18.13  & 13.43 \\
ShapeDiameter & 35.18  & 19.80 \\
SAMesh & 32.35 & 21.33 \\
Ours  & \textbf{61.78} & \textbf{70.54}  \\
\bottomrule
\end{tabular}
\end{table}

\noindent\textit{Quantitative Evaluation.}
We compute two metrics for all mesh vertices to assess the quality of the segmentation results: i) MUCoV, which quantifies the overall region-wise overlap between the predicted segments and the ground truth; and ii) Recall$_{50}$, which measures the which measures the fraction of ground-truth instances correctly identified with an IoU threshold of 50\%~\cite{ren2017end,lin2014microsoft}. 
The statistics are reported in Table~\ref{tab:quantitative-comparison-model}.
The segmentation results of our method achieve the best MUCoV and $\text{Recall}_{50}$ scores. These results confirm that our target problem is challenging, and validate the effectiveness of our proposed method in segmenting geometric texture instances over 3D models.

\section{Conclusion}
\label{sec:conclusion}

Existing model segmentation methods focus predominantly on semantics or low-level geometry.
In this work, we presented a method for the automatic recognition and segmentation of geometric textures on 3D surface models.
We first constructed a specialized benchmark for geometric texture segmentation. 
Then, we proposed a geometric texture-oriented segmentation method that follows a 2D-to-3D pipeline. By leveraging a fine-tuned visual segmentation model and DINO feature-based clustering, our method effectively identifies repetitive texture patterns across multiple views. Furthermore, we designed a weighted voting fusion strategy to achieve precise 3D segmentation. Building upon this, we developed a tool that enables users to segment geometric texture instances and to edit or retrieve models based on the segmentation result. 
Experimental results demonstrate the effectiveness of our method.

\begin{figure}[h]
    \centering
    \includegraphics[width=.85\linewidth]{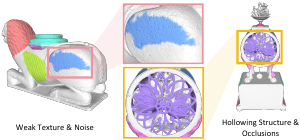}
    \caption{Two failure cases of our method.}
    \label{fig:limitations}
\end{figure}

\noindent\textit{Limitations and Future Works.}
However, our method fails in segmenting weak relief, hollow structures or heavy occlusions, as shown in Fig.~\ref{fig:limitations}. 
In the future, we plan to enrich our benchmark with more complex synthetic models and models reconstructed from real objects. 
Other future directions include: speeding up the segmentation process, integrating generative models for texture synthesis and transfer, and enabling interactive model segmentation~\cite{lang2024iseg,zhou2025splatmesh}.

\section*{Acknowledgements}
This work was supported in part by NSFC (62302313), ICFCRT (W2441020), GD Basic and Applied Basic Research Foundation (2023B1515120026), Shenzhen Science and Technology Program (KQTD20210811090044003, KJZD20240903100022028), and Scientific Development Fund from Guangdong Provincial Key Laboratory of Visual Media and Multidimensional Intelligence.

\bibliographystyle{eg-alpha-doi} 
\bibliography{GeoTex}

\newcommand{\etalchar}[1]{$^{#1}$}
\begin{thebibliography}{\uppercase{BWHSH25}}

\bibitem[BG05]{borg2005modern}
\textsc{Borg I., Groenen P.~J.}:
\newblock \emph{Modern multidimensional scaling: Theory and applications}.
\newblock Springer, 2005.

\bibitem[BIT04]{bhat2004geometric}
\textsc{Bhat P., Ingram S., Turk G.}:
\newblock Geometric texture synthesis by example.
\newblock In \emph{Proceedings of the 2004 Eurographics/ACM SIGGRAPH symposium
  on Geometry processing} (2004), pp.~41--44.

\bibitem[BTB{\etalchar{*}}18]{biasotti2018shrec}
\textsc{Biasotti S., Thompson E.~M., Barthe L., Berretti S., Giachetti A.,
  Lejemble Thibault e.~a.}:
\newblock Shrec'18 track: Recognition of geometric patterns over 3d models.
\newblock In \emph{Eurogr. Workshop 3D Object Retr.} (2018), The Eurographics
  Association.

\bibitem[BWHSH25]{binninger2025tetweave}
\textsc{Binninger A., Wiersma R., Herholz P., Sorkine-Hornung O.}:
\newblock Tetweave: Isosurface extraction using on-the-fly delaunay tetrahedral
  grids for gradient-based mesh optimization.
\newblock \emph{ACM Transactions on Graphics (TOG) 44}, 4 (2025), 1--19.

\bibitem[CFG{\etalchar{*}}15]{chang2015shapenet}
\textsc{Chang A.~X., Funkhouser T., Guibas L., et~al.}:
\newblock Shapenet: An information-rich 3d model repository.
\newblock \emph{arXiv preprint arXiv:1512.03012} (2015).

\bibitem[CGH{\etalchar{*}}25]{carion2025sam3}
\textsc{Carion N., Gustafson L., Hu Y.-T., Debnath S., Hu R., Suris Didac
  e.~a.}:
\newblock Sam 3: Segment anything with concepts.
\newblock \emph{arXiv preprint arXiv:2511.16719} (2025).

\bibitem[CLZ{\etalchar{*}}25]{Chen2025CLIP}
\textsc{Chen H.-Y., Lai Z., Zhang H., Wang X., Eichner M., You Keen e.~a.}:
\newblock Contrastive localized language-image pre-training.
\newblock In \emph{Int. Conf. Learn. Represent. ICLR} (2025).

\bibitem[CMK{\etalchar{*}}14]{cimpoi2014describing}
\textsc{Cimpoi M., Maji S., Kokkinos I., Mohamed S., Vedaldi A.}:
\newblock Describing textures in the wild.
\newblock In \emph{{Proc. IEEE Comput. Soc. Conf. Comput. Vis. Pattern
  Recognit}} (2014), pp.~3606--3613.

\bibitem[CMS{\etalchar{*}}22]{cheng2022mask2former}
\textsc{Cheng B., Misra I., Schwing A.~G., Kirillov A., Girdhar R.}:
\newblock Masked-attention mask transformer for universal image segmentation.
\newblock In \emph{{Proc. IEEE Comput. Soc. Conf. Comput. Vis. Pattern
  Recognit}} (2022), pp.~1290--1299.

\bibitem[CMT{\etalchar{*}}26]{cohen2026decoupling}
\textsc{Cohen I., Meivar B., Tu P., Avidan S., Oren G.}:
\newblock Decoupling shape and texture in sam-2 via controlled texture
  replacement.
\newblock In \emph{{IEEE} Winter Conf. Appl. Comput. Vis.} (2026),
  pp.~4077--4086.

\bibitem[CSK21]{cheng2021maskformer}
\textsc{Cheng B., Schwing A.~G., Kirillov A.}:
\newblock Per-pixel classification is not all you need for semantic
  segmentation.
\newblock In \emph{Adv. Neural Inf. Process. Syst.} (2021).

\bibitem[DHM{\etalchar{*}}25]{duggal2025eval3d}
\textsc{Duggal S., Hu Y., Michel O., Kembhavi A., Freeman W.~T., Smith Noah~A
  e.~a.}:
\newblock Eval3d: Interpretable and fine-grained evaluation for 3d generation.
\newblock In \emph{{Proc. IEEE Comput. Soc. Conf. Comput. Vis. Pattern
  Recognit}} (2025), pp.~13326--13336.

\bibitem[Elb05]{elber2005geometric}
\textsc{Elber G.}:
\newblock Geometric texture modeling.
\newblock \emph{IEEE Computer Graphics and Applications 25}, 4 (2005), 66--76.

\bibitem[GDJ{\etalchar{*}}24]{ganapathi20243d}
\textsc{Ganapathi I.~I., Dharejo F.~A., Javed S., Ali S.~S., Werghi N.}:
\newblock 3d-texseg: unsupervised segmentation of 3d texture using mutual
  transformer learning.
\newblock In \emph{International Conference on 3D Vision (3DV)} (2024), IEEE,
  pp.~506--515.

\bibitem[GDJ{\etalchar{*}}25]{Ganapathi20253dTexSeg}
\textsc{Ganapathi I.~I., Dharejo F.~A., Javed S., Ali S.~S., Werghi N.}:
\newblock Unsupervised dual transformer learning for 3-d textured surface
  segmentation.
\newblock \emph{IEEE Trans. Neural Netw. Learn. Syst. 36}, 3 (2025),
  5020--5031.

\bibitem[GKOM18]{Guerrero2018PCPNet}
\textsc{Guerrero P., Kleiman Y., Ovsjanikov M., Mitra N.~J.}:
\newblock Pcpnet: Learning local shape properties from raw point clouds.
\newblock \emph{Computer Graphics Forum 37}, 2 (2018), 115--126.

\bibitem[HGDG17]{he2017maskrcnn}
\textsc{He K., Gkioxari G., Doll{\'a}r P., Girshick R.}:
\newblock Mask {R-CNN}.
\newblock In \emph{{Proc. IEEE Int. Conf. Comput. Vis.}} (2017),
  pp.~2961--2969.

\bibitem[HHF{\etalchar{*}}19]{hanocka2019meshcnn}
\textsc{Hanocka R., Hertz A., Fish N., Giryes R., Fleishman S., Cohen-Or D.}:
\newblock Meshcnn: A network with an edge.
\newblock \emph{ACM Trans. Graph. 38}, 4 (2019), 1--12.

\bibitem[HHGCO20]{hertz2020deep}
\textsc{Hertz A., Hanocka R., Giryes R., Cohen-Or D.}:
\newblock Deep geometric texture synthesis.
\newblock \emph{ACM Transactions on Graphics (TOG) 39}, 4 (2020), 108--1.

\bibitem[HSW{\etalchar{*}}22]{hu2022lora}
\textsc{Hu E.~J., Shen Y., Wallis P., Allen-Zhu Z., Li Y., Wang Shean e.~a.}:
\newblock {LoRA}: Low-rank adaptation of large language models.
\newblock In \emph{Int. Conf. Learn. Represent. ICLR} (2022), pp.~12513--12525.

\bibitem[KMR{\etalchar{*}}23]{Kirillov2023SAM}
\textsc{Kirillov A., Mintun E., Ravi N., Mao H., Rolland C., Gustafson Laura
  e.~a.}:
\newblock Segment anything.
\newblock In \emph{{Proc. IEEE Int. Conf. Comput. Vis.}} (2023),
  pp.~4015--4026.

\bibitem[KT03]{katz2003hierarchical}
\textsc{Katz S., Tal A.}:
\newblock Hierarchical mesh decomposition using fuzzy clustering and cuts.
\newblock \emph{ACM Trans. Graph. 22}, 3 (2003), 954--961.

\bibitem[LHGM]{lai2005geometric}
\textsc{Lai Y.-K., Hu S.-M., Gu D., Martin R.~R.}:
\newblock Geometric texture synthesis and transfer via geometry images.
\newblock In \emph{Proceedings of the 2005 ACM symposium on Solid and physical
  modeling}, pp.~15--26.

\bibitem[LMB{\etalchar{*}}14]{lin2014microsoft}
\textsc{Lin T.-Y., Maire M., Belongie S., Hays J., Perona P., Ramanan D.,
  Doll{\'a}r P., Zitnick C.~L.}:
\newblock Microsoft coco: Common objects in context.
\newblock In \emph{European conference on computer vision} (2014), Springer,
  pp.~740--755.

\bibitem[LSD15]{long2015fcn}
\textsc{Long J., Shelhamer E., Darrell T.}:
\newblock Fully convolutional networks for semantic segmentation.
\newblock In \emph{{Proc. IEEE Comput. Soc. Conf. Comput. Vis. Pattern
  Recognit}} (2015), pp.~3431--3440.

\bibitem[LUX{\etalchar{*}}25]{liu2025partfield}
\textsc{Liu M., Uy M.~A., Xiang D., Su H., Fidler S., Sharp N., Gao J.}:
\newblock Partfield: Learning 3d feature fields for part segmentation and
  beyond.
\newblock In \emph{{Proc. IEEE Int. Conf. Comput. Vis.}} (2025), IEEE,
  pp.~9704--9715.

\bibitem[LXD{\etalchar{*}}24]{lang2024iseg}
\textsc{Lang I., Xu F., Decatur D., Babu S., Hanocka R.}:
\newblock iseg: Interactive 3d segmentation via interactive attention.
\newblock In \emph{SIGGRAPH Asia} (2024), pp.~1--11.

\bibitem[LYW{\etalchar{*}}20]{liang2020survey}
\textsc{Liang C., Yin J., Wu J., Wang J., Wei M., Guo Y.}:
\newblock A survey of 3d mesh segmentation based on clustering analysis.
\newblock \emph{Journal of Computer-Aided Design \& Computer Graphics 32}, 4
  (2020), 680--692.

\bibitem[MWS{\etalchar{*}}25]{meng2025text2vdm}
\textsc{Meng H., Wang D., Shao Z., Liu L., Wang Z.}:
\newblock Text2vdm: Text to vector displacement maps for expressive and
  interactive 3d sculpting.
\newblock In \emph{{Proc. IEEE Int. Conf. Comput. Vis.}} (2025),
  pp.~16882--16892.

\bibitem[MZC{\etalchar{*}}19]{Mo2019PartNet}
\textsc{Mo K., Zhu S., Chang A.~X., Yi L., Tripathi S., Guibas Leonidas~J.
  e.~a.}:
\newblock Partnet: A large-scale benchmark for fine-grained and hierarchical
  part-level 3d object understanding.
\newblock In \emph{{Proc. IEEE Comput. Soc. Conf. Comput. Vis. Pattern
  Recognit}} (2019), pp.~909--918.

\bibitem[ODM{\etalchar{*}}24]{oquab2024dinov2}
\textsc{Oquab M., Darcet T., Moutakanni T., Vo H.~V., Szafraniec M., Khalidov
  Vasil e.~a.}:
\newblock {DINOv2}: Learning robust visual features without supervision.
\newblock \emph{Transactions on Machine Learning Research} (2024).

\bibitem[PTB24]{paolini2024binary}
\textsc{Paolini G., Tortorici C., Berretti S.}:
\newblock Binary segmentation of relief patterns on point clouds.
\newblock \emph{Computers \& Graphics 123} (2024), 104020.

\bibitem[PTB{\etalchar{*}}25]{paolini2025shrec}
\textsc{Paolini G., Tortorici C., Berretti S., et~al.}:
\newblock Shrec'25 track: Retrieval and segmentation of multiple relief
  patterns.
\newblock In \emph{Eurographics Workshop on 3D Object Retrieval} (2025),
  pp.~1--8.

\bibitem[QSMG17]{qi2017pointnet}
\textsc{Qi C.~R., Su H., Mo K., Guibas L.~J.}:
\newblock {PointNet}: Deep learning on point sets for 3d classification and
  segmentation.
\newblock In \emph{{Proc. IEEE Comput. Soc. Conf. Comput. Vis. Pattern
  Recognit}} (2017), pp.~652--660.

\bibitem[QYSG17]{qi2017pointnetplusplus}
\textsc{Qi C.~R., Yi L., Su H., Guibas L.~J.}:
\newblock {PointNet++}: Deep hierarchical feature learning on point sets in a
  metric space.
\newblock In \emph{Adv. Neural Inf. Process. Syst.} (2017).

\bibitem[RFB15]{ronneberger2015unet}
\textsc{Ronneberger O., Fischer P., Brox T.}:
\newblock {U-Net}: Convolutional networks for biomedical image segmentation.
\newblock In \emph{Med. Image Comput. Comput. Assist. Interv.} (2015),
  pp.~234--241.

\bibitem[RZ17]{ren2017end}
\textsc{Ren M., Zemel R.~S.}:
\newblock End-to-end instance segmentation with recurrent attention.
\newblock In \emph{IEEE Conference on Computer Vision and Pattern Recognition
  (CVPR)} (2017), pp.~293--301.

\bibitem[Sha08]{shamir2008survey}
\textsc{Shamir A.}:
\newblock A survey on mesh segmentation techniques.
\newblock In \emph{Comp. Graph. Forum} (2008), vol.~27, pp.~1539--1556.

\bibitem[SSCO08]{shapira2008consistent}
\textsc{Shapira L., Shamir A., Cohen-Or D.}:
\newblock Consistent mesh partitioning and skeletonisation using the shape
  diameter function.
\newblock \emph{The Visual Computer 24}, 4 (2008), 249--259.

\bibitem[TBOW21]{tortorici2021convolution}
\textsc{Tortorici C., Berretti S., Obeid A., Werghi N.}:
\newblock Convolution operations for relief-pattern retrieval, segmentation and
  classification on mesh manifolds.
\newblock \emph{Pattern Recognition Letters 142} (2021), 32--38.

\bibitem[TCC{\etalchar{*}}26]{sam3dteam2026sam3d}
\textsc{Team S.~D., Chen X., Chu F.-J., Gleize P., Liang K.~J., Alexander~Sax
  e.~a.}:
\newblock Sam 3d: 3dfy anything in images.
\newblock \emph{arxiv preprint arxiv2511.16624} (2026).

\bibitem[TZF{\etalchar{*}}24a]{tang2024segment}
\textsc{Tang G., Zhao W., Ford L., Benhaim D., Zhang P.}:
\newblock Segment any mesh.
\newblock \emph{arXiv preprint arXiv:2408.13679} (2024).

\bibitem[TZF{\etalchar{*}}24b]{Tang2024SAM}
\textsc{Tang G., Zhao W., Ford L., Benhaim D., Zhang P.}:
\newblock Segment any mesh: Zero-shot mesh part segmentation via lifting
  segment anything 2 to 3d.
\newblock \emph{arXiv preprint arxiv:2408.13679} (2024).

\bibitem[VPP{\etalchar{*}}23]{Vecchio2023Met}
\textsc{Vecchio G., Prezzavento L., Pino C., Rundo F., Palazzo S., Spampinato
  C.}:
\newblock Met: A graph transformer for semantic segmentation of 3d meshes.
\newblock \emph{Comput. Vis. Image Underst. 235}, C (2023).

\bibitem[WGS{\etalchar{*}}18]{wang2018shape}
\textsc{Wang P., Gan Y., Shui P., Yu F., Zhang Y., Chen Songle e.~a.}:
\newblock 3d shape segmentation via shape fully convolutional networks.
\newblock \emph{Computers \& Graphics 70} (2018), 128--139.

\bibitem[WSL{\etalchar{*}}19]{Wang2019DGCNN}
\textsc{Wang Y., Sun Y., Liu Z., Sarma S.~E., Bronstein M.~M., Solomon J.~M.}:
\newblock Dynamic graph cnn for learning on point clouds.
\newblock \emph{ACM Trans. Graph. 38} (2019).

\bibitem[XHW{\etalchar{*}}26]{xiong2026paver}
\textsc{Xiong W., Hu Z., Wu Y., Song P., Zheng J.}:
\newblock Paver: Element-based pattern creation on 3d free-form surfaces.
\newblock \emph{Computer-Aided Design} (2026), 104031.

\bibitem[XYQ{\etalchar{*}}25]{xu2025sampro3d}
\textsc{Xu M., Yin X., Qiu L., Liu Y., Tong X., Han X.}:
\newblock Sampro3d: Locating sam prompts in 3d for zero-shot instance
  segmentation.
\newblock In \emph{International Conference on 3D Vision (3DV)} (2025), IEEE,
  pp.~1222--1232.

\bibitem[XZSJ18]{xiong2018shape}
\textsc{Xiong W., Zhang P., Sander P.~V., Joneja A.}:
\newblock Shape-inspired architectural design.
\newblock In \emph{Proc. ACM SIGGRAPH Interactive 3D Graphics and Games}
  (2018), pp.~1--10.

\bibitem[XZZ{\etalchar{*}}14]{xiong2014robust}
\textsc{Xiong S., Zhang J., Zheng J., Cai J., Liu L.}:
\newblock Robust surface reconstruction via dictionary learning.
\newblock \emph{ACM Transactions on Graphics (TOG) 33}, 6 (2014), 1--12.

\bibitem[YCK{\etalchar{*}}25]{yang2025genvdm}
\textsc{Yang Y., Chen Q., Kim V.~G., Chaudhuri S., Huang Q., Chen Z.}:
\newblock Genvdm: Generating vector displacement maps from a single image.
\newblock In \emph{{Proc. IEEE Comput. Soc. Conf. Comput. Vis. Pattern
  Recognit}} (2025), pp.~26618--26629.

\bibitem[YLX{\etalchar{*}}24]{yin2024sai3d}
\textsc{Yin Y., Liu Y., Xiao Y., Cohen-Or D., Huang J., Chen B.}:
\newblock Sai3d: Segment any instance in 3d scenes.
\newblock In \emph{{Proc. IEEE Comput. Soc. Conf. Comput. Vis. Pattern
  Recognit}} (2024), pp.~3292--3302.

\bibitem[YRWX25]{ying2025chord}
\textsc{Ying Z., Rong B., Wang J., Xu M.}:
\newblock Chord: Chain of rendering decomposition for pbr material estimation
  from generated texture images.
\newblock In \emph{Proc. SIGGRAPH Asia} (2025), ACM.

\bibitem[YZZW24]{yan2024maskclustering}
\textsc{Yan M., Zhang J., Zhu Y., Wang H.}:
\newblock Maskclustering: View consensus based mask graph clustering for
  open-vocabulary 3d instance segmentation.
\newblock In \emph{{Proc. IEEE Comput. Soc. Conf. Comput. Vis. Pattern
  Recognit}} (2024), pp.~28274--28284.

\bibitem[ZDZ{\etalchar{*}}15]{zhang2015guided}
\textsc{Zhang W., Deng B., Zhang J., Bouaziz S., Liu L.}:
\newblock Guided mesh normal filtering.
\newblock In \emph{Comp. Graph. Forum} (2015), vol.~34, pp.~23--34.

\bibitem[ZHC{\etalchar{*}}25]{zhou2025splatmesh}
\textsc{Zhou K., Hong L., Chang X., Zhong Y., Xie E., Dong H., Li Z., Yang Y.,
  Li Z., Zhang W.}:
\newblock Splatmesh: Interactive 3d segmentation and editing using mesh-based
  gaussian splatting.
\newblock In \emph{IEEE/CVF Conference on Computer Vision and Pattern
  Recognition Workshops (CVPRW)} (2025), pp.~01--12.

\bibitem[ZHR{\etalchar{*}}26]{ZHOU2026OV3dSeg}
\textsc{Zhou J., Huang X., Ren Y., Fan D., Gu S., Bin~Tian e.~a.}:
\newblock Ov3dseg-vggt: Open-vocabulary 3d segmentation with visual
  geometry-grounded transformers.
\newblock \emph{Visual Informatics} (2026), 100311.

\bibitem[ZJ16]{Zhou2016Thingi10K}
\textsc{Zhou Q., Jacobson A.}:
\newblock Thingi10k: A dataset of 10,000 3d-printing models.
\newblock \emph{arXiv preprint arXiv:1605.04797} (2016).

\bibitem[ZXL{\etalchar{*}}24]{zhong2024meshsegmenter}
\textsc{Zhong Z., Xu Y., Li J., Xu J., Li Z., Yu Chaohui e.~a.}:
\newblock Meshsegmenter: Zero-shot mesh semantic segmentation via texture
  synthesis.
\newblock In \emph{{Eur. Conf. Comput. Vis.}} (2024), pp.~182--199.

\bibitem[ZXZD24]{zheng2024design}
\textsc{Zheng J., XIONG W., Zeng J., Davis E.~D.}:
\newblock Design element placement on 3d surfaces, July~11 2024.
\newblock US Patent App. 18/279,130.

\end{thebibliography}




\end{document}